\documentclass[aps,prl,twocolumn,superscriptaddress]{revtex4}

\usepackage[dvips]{graphics, color}

\input epsf

\begin{document}

\newcommand{\Ra}{$^{85}$Rb }
\newcommand{\Rb}{$^{87}$Rb }
\newcommand{\ra}{$^{85}$Rb}
\newcommand{\rb}{$^{87}$Rb}

\title{Studying a dual-species BEC with tunable interactions}
\author{S. B. Papp}
\email[Current address: Norman Bridge Laboratory of Physics 12-33, California Institute of Technology, Pasadena, CA 91125, USA]{}
\address{JILA, National Institute of Standards and Technology
and University of Colorado, Boulder, Colorado 80309-0440, USA}
\author{J. M. Pino}
\address{JILA, National Institute of Standards and Technology
and University of Colorado, Boulder, Colorado 80309-0440, USA}
\author{C. E. Wieman}
\address{University of British Columbia, Vancouver, BC V6T 1Z1, CANADA}
\address{JILA, National Institute of Standards and Technology
and University of Colorado, Boulder, Colorado 80309-0440, USA}
\date{\today}

\begin{abstract}

We report on the observation of controllable spatial separation in a dual-species Bose-Einstein condensate (BEC) with \Ra and \rb.  Interparticle interactions between the different components can change the miscibility of the two quantum fluids.  In our experiments, we clearly observe the immiscible nature of the two simultaneously Bose-condensed species via their spatial separation.  Furthermore the \Ra Feshbach resonance near 155 G is used to change them between miscible and immiscible by tuning the \Ra scattering length.  Our apparatus is also able to create \Ra condensates with up to $8\times10^4$ atoms which represents a significant improvement over previous work.

\end{abstract}
\maketitle

Two-component quantum fluids have long been known to exhibit rich physics that is not accessible in a simple single-component fluid \cite{Guttman1953a}.  The miscibility of two quantum fluids represents a fundamental property of a two-component system, and it is partially controlled by interparticle interactions.  Building on the achievement of dilute-gas Bose-Einstein condensation \cite{Dalfovo1999a}, mixtures of ultracold quantum gases provide a unique system for the study of interacting multi-component quantum fluids.  Moreover resonant control of two-body interactions in ultracold gases is possible via magnetic-field Feshbach resonances \cite{Kohler2006a} and provides a parameter which directly alters the miscibility of the dual-species BEC.

The first two-component condensate was produced with different hyperfine states of $^{87}$Rb, and evidence for repulsive interactions between the hyperfine states was reported \cite{Myatt1997a}.  Further work with different hyperfine states of \Rb studied the dynamics of interpenetrating quantum fluids \cite{Hall1998a}.  In spinor condensates of Na both miscible and immiscible two-component condensates were observed for the first time \cite{Stenger1999a}.  In further work, long-lived metastable excited states were realized with spinor Na condensates \cite{Miesner1999a,Stamper-Kurn1999b}.  More recently a dual-species BEC of $^{41}$K and $^{87}$Rb was reported in Ref. \cite{Modugno2002a} where scissors-like collective oscillations were observed as a result of off-axis collisions.

Two-component quantum gases have also long been the subject of significant theoretical interest; topics of study include the density pattern and phase separation of the components \cite{Ho1996a,Ao1998a,Esry1997a,Timmermans1998a,Pu1998a,Barankov2002a} and the spectrum of collective excitations \cite{Busch1997a,Graham1998b}.  Spatial separation of two immiscible quantum fluids in a trap is typified by a ball-and-shell ground-state structure in which one fluid forms a low density shell around the other.  This structure depends on the relative strength of interspecies and single-species interactions \cite{Ho1996a}.  Other ground-state structures are expected as well and a general classification of all possible ground-state spatial structures has been presented in Ref. \cite{Trippenbach2000c}.  A mixture of \Ra and \Rb has long been considered a promising experimental system for studying two interpenetrating quantum fluids, and a route to dual-species BEC with Rb was proposed in Ref. \cite{Burke1998a}.  The tendency for spatial separation to occur in a \Ra--\Rb mixture can be predicted by applying the theoretical analysis of Refs. \cite{Riboli2002a,Jezek2002a} for the Thomas-Fermi limit.  Based on their analysis we define the parameter
\begin{equation}
\Delta = \frac{a_{85}\,a_{87}}{a_{85-87}^2} - 1
\end{equation}
that depends upon the ratio of the single species and interspecies interactions.  In the case where $a_{85-87}$ is positive two regimes emerge: $\Delta>0$ where the two condensates are miscible and $\Delta<0$ where they are immiscible due to the strong interspecies repulsion.  For simplicity we have neglected the small difference in mass of \Ra and \Rb here.

In this Letter we report the production of a dual-species BEC of \Ra and \Rb with tunable interactions.  Immiscibility of the two-component quantum gas is observed as a dramatic departure in the density distribution from that of the symmetry of the external trapping potential.  Surprisingly we observe the robust formation of multiple, non-overlapped, single-species BEC ``cloudlets'' which represent an interesting metastable excited state. We find that the degree of spatial separation between the two species can be controlled with a tunable \Ra scattering length.

\begin{figure}[htbp]
\begin{center}
\scalebox{0.6}[0.6]{\includegraphics{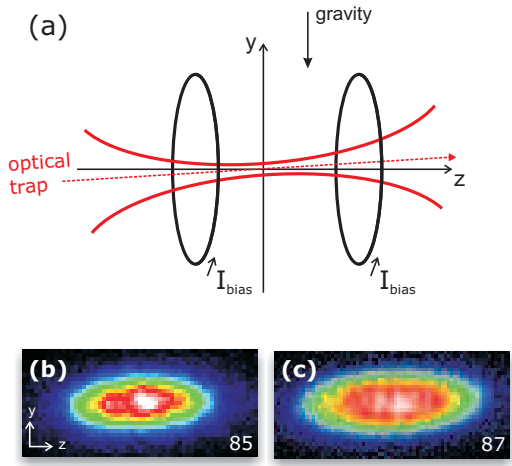}}
\caption{(a) Schematic drawing of the optical trap laser beam and the magnetic-field coils.  The curvature of the magnetic field enhances confinement along the axial direction (z) of the optical trap.  The center position of the trapping potential along its axis is influenced by the tilt of the optical trap relative to gravity, and the relative position of the optical trap beam focus and the coils.  Optical trap tilt and misalignment of the beam focus and the coils have been exaggerated for clarity.  Absorption images are acquired along the axis which is into the page.  (b, c)  Absorption images of single-species \Ra and \Rb condensates which demonstrate the smooth single-species density profiles consistent with the shape of the trapping potential.  Prior to absorption imaging, the gas is released from the optical trap and it expands for 20 ms; the magnetic field remains on for the first 10 ms of expansion to prevent collapse of the \Ra gas \cite{Roberts2001a, Donley2001a}. \label{apparatus}}
\end{center}
\end{figure}

Our apparatus and evaporative cooling routine are described in Ref. \cite{Papp2006a}.  Atoms are initially collected from a vapor into a two-species magneto-optical trap (MOT).  After magnetic transfer to another region of the apparatus with lower vacuum pressure the atoms are prepared into the $^{85}$Rb $|f=2, m_f=-2\rangle$ state and $^{87}$Rb $|f=1, m_f=-1\rangle$ state and loaded into a magnetic trap.  Selective rf evaporation is used to lower the temperature of the $^{87}$Rb gas to 10 $\mu$K, while the $^{85}$Rb gas is sympathetically cooled through thermal contact with the $^{87}$Rb \cite{Bloch2001a}.  The two-species gas is then loaded into an optical dipole trap [Fig. \ref{apparatus}], while maintaining a large, nearly uniform magnetic field. Further evaporation to BEC is performed by lowering the depth of the optical potential.  Unless otherwise noted, after evaporation the depth of the optical trap is raised in 50 ms to  $\sim1\mu$K with a measured \Ra radial (axial) trap frequency of $2\pi\times130$ Hz ($2\pi\times2.9$ Hz).  While the \Rb radial frequency is nearly identical to that of \ra, the axial frequency ($2\pi\times2.6$ Hz) is smaller due to a different magnetic moment and the curvature of the magnetic field.

To create a \Ra BEC it is important to use a weakly confining optical trap to reduce density-dependent inelastic loss and to control the scattering length of \Ra using the Feshbach resonance.  Specifically, condensates are created only in a small range of magnetic field near 163.1 G where the \Ra interaction is repulsive \cite{Cornish2000a,Claussen2003a} and inelastic collision rates are sufficiently small.  By optimizing the evaporation rate and the precise number ratio of the two species at the start of evaporation, we are able to create a single-species \Ra BEC with up to $8\times10^4$ atoms, a \Rb BEC with $2\times10^6$ atoms, or a BEC of both species.  Absorption images of single-species condensates after expansion are shown in Fig. \ref{apparatus}.  The elongated axial direction (z) of all the absorption images in this paper show the spatial density of the gas since the trapping frequency is small. The images show the momentum density in the radial direction (y).  Several interspecies Feshbach resonances are accessible in our system including the $s$-wave resonances used in Ref. \cite{Papp2006a} as well as recently discovered $p$-wave resonances at 257.8 G and $\approx330$ G \cite{Ticknor2004a}.  For the magnetic fields used in this work, however, the interspecies scattering length is  +213(7) $a_0$ \cite{Burke1999a} and the \Rb scattering length is +99 $a_0$ \cite{Kempen2002a}.

\begin{figure}[htbp]
\begin{center}
\scalebox{0.6}[0.6]{\includegraphics{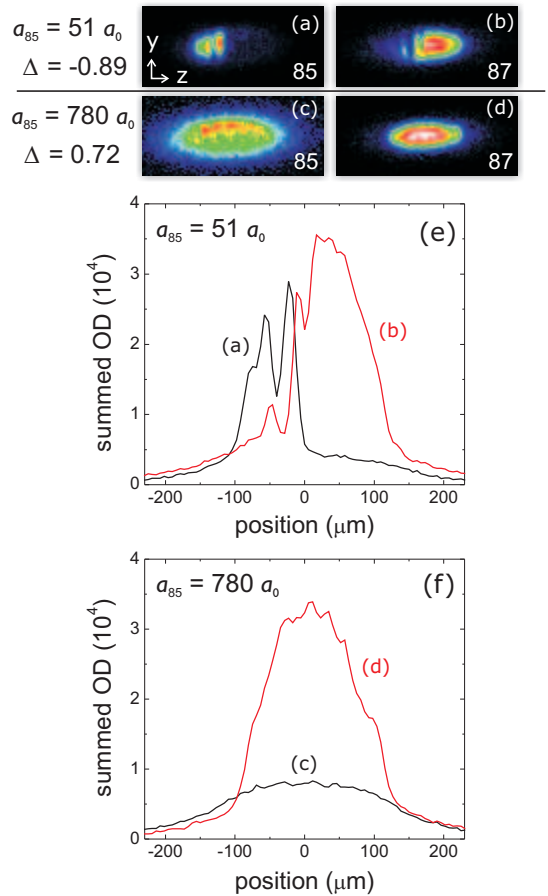}}
\caption{Absorption images at two different \Ra scattering lengths that display immiscible (a,b) and miscible (c,d) behavior.  Each image in the pairs (a,b) and (c,d) were acquired with different condensates after creation under the same conditions.  The sum of each image pair should be interpreted as the total density of the expanded gas.   (e, f)  The measured two-dimensional optical density (OD) was summed in the remaining radial direction to produce a cross section.  The \Ra scattering length was 51 $a_0$ for images (a, b) and it was 780 $a_0$ for images (c, d).  The number of \Ra (\rb) atoms is approximately $4\times10^4$ ($9\times10^4$).     \label{cross}}
\end{center}
\end{figure}

Interesting physics occurs when we create a simultaneously Bose-condensed sample of \Ra and \rb. The dual BEC exhibits either miscible or immiscible behavior, depending on the magnetic field, as seen in Fig. \ref{cross}.  After preparation of a dual species BEC, the magnetic field was swept in 400 ms from 163.1 G to a value in the range of 164.6 G to 158.6 G; the sweep covered a range in $\Delta$ of approximately -1 to +1 (50 $a_0$ to 900 $a_0$).  The rate of this sweep was measured to be slow with respect to the response of the gases in the radial and axial directions of the trap.   Absorption images of each gas were acquired after the sweep. The images  are shown for $\Delta = -0.89$ [Fig. \ref{cross} (a) and (b)] and $\Delta = 0.72$ [Fig. \ref{cross} (c) and (d)].  We observe a dramatic spatial separation of the two gases when $\Delta$ is negative.  The irregular spikes in the density patterns of the two condensates will be discussed in more detail below.  When $\Delta$ is positive, the coexistence of each species throughout the axial direction is observed and the sharp density spikes of the immiscible condensates give way to a smooth overall density.  Spatial separation was observed to return when the magnetic-field sweep to $\Delta>0$ was immediately followed by a second opposite field sweep to $\Delta<0$.  The penetration distance in the axial direction over which the two expanded gases overlap is estimated to be approximately 30 $\mu$m when $\Delta = -0.89$.

\begin{figure}[htbp]
\begin{center}
\scalebox{1}[1]{\includegraphics{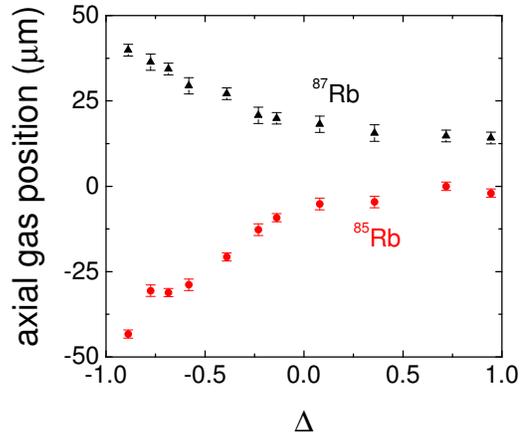}}
\caption{Measured axial position of the \Ra and \Rb gases as a function of the parameter $\Delta$.  We report the uncertainty in the position as the shot-to-shot reproducibility of our measurements.  When $\Delta>0$ the measured difference in the position of each gas is in good agreement with that expected for single-species condensates.   \label{position}}
\end{center}
\end{figure}

To characterize the degree of spatial separation as $\Delta$ is varied, we have extracted the axial center-of-mass location of each species from images similar to those shown in Fig. \ref{cross}.  In this experiment an offset between the axial trap centers for each species was intentionally introduced [Fig. \ref{apparatus} (a)] to reflect the spatial separation in the center-of-mass location.  The offset causes the two immiscible quantum gases to separate primarily in the axial direction with each species preferring a well-defined side of the trap.  The results of this experiment are plotted in Fig. \ref{position} as a function of $\Delta$.  The centers of each BEC gradually come together as $\Delta$ is increased from -0.9 toward zero.  For $\Delta>0$ there is little change as the two species can then coexist throughout the trap. However, residual gravitational sag likely induces incomplete spatial overlap even when $\Delta>0$ \cite{Riboli2002a}.

\begin{figure}[htbp]
\begin{center}
\scalebox{0.6}[0.6]{\includegraphics{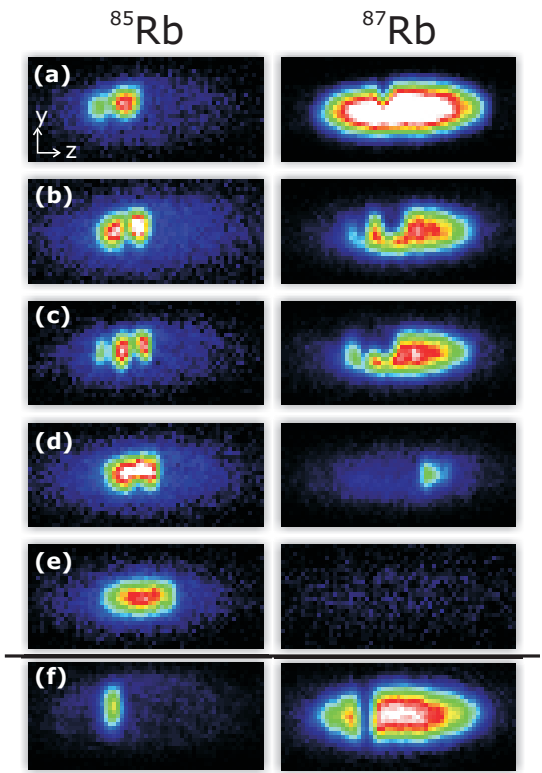}}
\caption{(a-e) Absorption images of \Ra (left column) and \Rb (right column) condensates demonstrating the immiscibility of the quantum gases.  For clarity the OD of each \Rb image was scaled by a factor of 0.5 and the size of each image is 330 $\mu$m $\times$ 150 $\mu$m.  The number of \Ra particles varies from approximately $1.2\times10^4$ in (a) to $2.2\times10^4$ in (e); the \Rb number varies from $1.4\times10^5$ in (a) to $3\times10^3$ in (e).  For comparison absorption images of single species \Ra and \Rb condensates are shown in Fig. \ref{apparatus} (b, c).  (f) Absorption images following recompression of the optical trap that demonstrate a symmetrical radial distribution.  \label{cloud}}
\end{center}
\end{figure}

We have studied the different immiscible density patterns as a function of the number of each Rb species.  We vary the relative number in the two-species MOT to change the number of atoms in the BECs.  Figure \ref{cloud} shows absorption images of both \Ra and \Rb condensates immediately after evaporation to a fixed final trap depth of $\approx$100 nK, and  $\Delta = -0.82$ (81 $a_0$) so that the two quantum fluids are always immiscible.  In each image a different condensate is shown since the measurement process is destructive.  The trap was not recompressed in order to accentuate the spatial separation of the two species.

When a condensate of each species is present in the trap we observe ``holes'' in the \Rb image that correspond to ``spikes'' at approximately the same location in the \Ra image [Fig. \ref{cloud} (a)-(d)].  This behavior is consistent with spatial separation of two immiscible quantum fluids.  However the images are not consistent with any known phase-separation structure expected for the ground state of a two-component system as described in Ref. \cite{Trippenbach2000c}.  The \Ra gas is observed to split into multiple separated cloudlets that appear as distinct holes in the \Rb gas.  In Fig. \ref{cloud} (a) and (b) two \Ra cloudlets are observed, and three cloudlets are observed in image (c).  These images represent long-lived presumably metastable excited states of the two-component system that depend on the relative number of each species.  If the initial number of \Ra atoms at the start of evaporation is too large, the \Rb gas cannot be cooled to condensation and remains a dilute thermal gas; in this case [Fig. \ref{cloud} (e)] the \Ra condensate shape is unperturbed and well approximated by a Thomas-Fermi model.

The immiscible gases shown in Fig. \ref{cloud} are not symmetric with respect to either the radial or axial directions of the trap.  These asymmetries are caused by gravitational sag of the trap and can be reduced by recompression to increase the trap depth [Fig. \ref{cloud} (f)].  Symmetry in the radial direction is restored  but spatial separation is still clearly observed along the axial dimension.  Similarly we have verified that offsets in the center-of-mass axial position of each species can be controlled via precise alignment of the optical trap focus with respect to the magnetic field [Fig. \ref{apparatus} (a)].  Although the precise shape, size, and position of the immiscible gases is affected by details of the trapping potential, the observation of spatial separation and the formation of separated \Ra cloudlets is repeatable under a variety of experimental conditions.  The spatial separation and cloudlet behavior are observed during the evaporation process as soon as the temperature of both species is lowered below the condensation temperature. Furthermore, the dual-species BEC remains spatially separated and individual cloudlets of \Ra do not coalesce during the 1 sec lifetime of the \Ra BEC, which includes many radial and a few axial trap oscillations.

We lack a detailed theoretical understanding of \Ra cloudlet formation in our experiments.  The geometry of our system is complicated by the elongated axial direction of the trap with an aspect ratio near 50 and the presence of gravitational sag.  In an analysis of the favorability for cloudlet formation the energy cost of forming and maintaining extra boundary surface area must be weighed against the cost of all particles residing in the same cloud \cite{Timmermans1998a}.  These questions appear to require the solution of coupled Gross-Pitaevskii equations for the dual-species BEC wave-functions which is beyond the scope of this paper \cite{Esry1997a}.  Recent theoretical simulations of our system have obtained qualitative agreement with the data presented here \cite{shai}.

In summary, we have produced a dual-species BEC of \Ra and \Rb with tunable interactions.  Immiscibility of the two quantum fluids was observed as a departure in the gas density pattern from that expected due to the symmetry of the potential.  By varying the \Ra scattering length the degree of spatial separation was characterized.  Surprisingly immiscible condensates are observed to form multiple spatially-separated cloudlets.  In the future, studies of the collective-excitation spectrum of a two-component BEC may be possible \cite{Graham1998b}.  Numerous Feshbach resonances are available in the \Ra--\Rb system making it possible to study a two-component condensate with a tunable interspecies interaction.  In addition, the dynamics of single-species condensate collapse may be modified by the presence of a second condensate component.

We gratefully acknowledge useful discussions with Mark Edwards, Shai Ronen, Deborah Jin, and Eric Cornell.  S.B.P acknowledges support from the NSF-GRFP.  This work has been supported by NSF and ONR.

\bibliographystyle{prsty}
\bibliography{../bib/sp_refs}

\end{document}